\documentclass[letterpaper]{article} 
\usepackage{aaai23}  
\usepackage{times}  
\usepackage{helvet}  
\usepackage{courier}  
\usepackage[hyphens]{url}  
\usepackage{graphicx} 
\usepackage{natbib}  
\usepackage{caption} 
\usepackage{subcaption}
\usepackage{syntonly}

\usepackage{algorithm}
\usepackage{algorithmic}
\usepackage{comment}
\usepackage{xcolor}
\usepackage{booktabs}
\usepackage{amsmath}
\usepackage{amsfonts}

\usepackage{bbding}
\usepackage{pifont}
\usepackage{wasysym}
\usepackage{amssymb}
\usepackage{tikz}
\usepackage{pgfplots}

\usepackage{newfloat}
\usepackage{listings}
\DeclareCaptionStyle{ruled}{labelfont=normalfont,labelsep=colon,strut=off} 
\floatstyle{ruled}
\newfloat{listing}{tb}{lst}{}
\floatname{listing}{Listing}
\title{Efficient ML Models for Practical Secure Inference}
\author {
    Vinod Ganesan\textsuperscript{\rm 1 \rm 2},
    Anwesh Bhattacharya\textsuperscript{\rm 1},
    Pratyush Kumar\textsuperscript{\rm 1 \rm, 2},
    Divya Gupta\textsuperscript{\rm 1},
    Rahul Sharma\textsuperscript{\rm 1},
    Nishanth Chandran\textsuperscript{\rm 1}
}
\affiliations {
    \textsuperscript{\rm 1} Microsoft Research, India\\
    \textsuperscript{\rm 2} IIT Madras\\
    \{t-viganesan, t-anweshb, pratykumar, divya.gupta, rahsha, nichandr\}@microsoft.com
}

\newcommand{\pk}[1]{{\color{violet} \small (PK: #1)}}
\newcommand{\ab}[1]{{\color{olive} \small (AB: #1)}}
\newcommand{\vg}[1]{{\color{cyan} \small (VG: #1)}}
\newcommand{\rs}[1]{{\color{orange} \small (RS: #1)}}
\newcommand{\dg}[1]{{\color{blue} \small (DG: #1)}}
\newcommand{\nc}[1]{{\color{purple} \small (NC: #1)}}

\newcommand{\shuffle}{shuffle operator}

\newcommand{\mult}{\#\mathsf{MULT}}
\newcommand{\cin}{C_{\mathsf{in}}}
\newcommand{\cout}{C_{\mathsf{out}}}

\definecolor{myblue}{HTML}{007BA7}
\definecolor{myred}{HTML}{800020}
\definecolor{mygreen}{HTML}{008000}
\definecolor{paretoblue}{HTML}{26619C}
\definecolor{paretogreen}{HTML}{087830}
\pgfplotsset{compat=1.17}

\begin{document}

\maketitle

\begin{abstract}
ML-as-a-service continues to grow, and so does the need for very strong privacy guarantees.
Secure inference has emerged as a potential solution, wherein cryptographic primitives allow inference without revealing users' inputs to a model provider or model's weights to a user.
For instance, the model provider could be a diagnostics company that has trained a state-of-the-art DenseNet-121 model for interpreting a chest X-ray and the user could be a patient at a hospital.
While secure inference is in principle feasible for this setting, there are no existing techniques that make it practical at scale.
The CrypTFlow2 framework provides a potential solution with its ability to automatically and correctly translate clear-text inference to secure inference for arbitrary models. 
However, the resultant secure inference from CrypTFlow2 is impractically expensive: Almost 3TB of communication is required to interpret a single X-ray on DenseNet-121. 

In this paper, we address this outstanding challenge of inefficiency of secure inference with three contributions. 
First, we show that the primary bottlenecks in secure inference are large linear layers which can be optimized with the choice of network backbone and the use of operators developed for efficient clear-text inference. 
This finding and emphasis deviates from many recent works which focus on optimizing non-linear activation layers when performing secure inference of smaller networks. 
Second, based on analysis of a bottle-necked convolution layer, we design a X-operator which is a more efficient drop-in replacement.
The X-operator combines various ideas which we found to be efficient for secure inference: factorizing linear operations along with using shuffle operations and larger proportion of additions, both of which are ``free'' (i.e., require no communication) in secure inference.
Third, we show that the fast Winograd convolution algorithm further improves efficiency of secure inference.
We add Winograd algorithm support to the Athos compiler frontend of CrypTFlow2, thereby automatically improving efficiency for most deep networks. 
In combination, these three optimizations prove to be highly effective for the problem of X-ray interpretation trained on the CheXpert dataset:
Relative to state-of-the-art models, we identify a model with about 8$\times$ lower communication cost with a negligible drop in AUC of 0.006, and another model with over 30$\times$ lower cost with a drop in AUC of 0.02.
We believe that inference with very strong privacy guarantees can be made practical with such efficient ML models. 
\end{abstract}

\section{Introduction}
\label{sec:intro}

\noindent 
Often large Machine Learning (ML) models are deployed on cloud computers which run inference on user data, a template loosely called ML-as-a-service (MLaaS). 
A growing concern with this template is the need for guaranteeing user privacy.
This is particularly important in domains such as healthcare where user privacy is paramount and often legally binding~\cite{hipaa, caring}.
In addition, ML models for healthcare, such as models for prognosis, have high commercial value given the challenges with obtaining training data and thus need to be protected.
One of the approaches to interface private data with protected models is \textit{secure inference} with cryptographically-secure 2-party computation (2PC) protocols.
Specifically, secure inference achieves the following: If a user has private data $x$ on which she wants to find the output of inference $M(w, x)$ on a model $M$ with weights $w$, then this should be possible, without any trusted intermediary, and without the user learning anything about $w$ (except for $M(w, x)$) or the model provider learning anything about $x$. 
Recent advancements in secure inference include \cite{secureml, gazelle, delphi, cryptflow2, 
cheetah}. 

With these advances, is secure inference for deep models practical today?
To ground a notion of what is practical, we take a specific example of chest X-ray interpretation. 
The largest public dataset of labelled chest X-rays is CheXpert \cite{chexpert} with over 200K X-rays of size 320x320 labelled with 14 labels of 3-classes.
Deep models such as DenseNet-121 with 121 layers and over 7.5M parameters are required to achieve state-of-the-art accuracy on this task, \emph{i.e.}, an AUC of around 0.888.
We could consider secure inference practical if a such a model trained on CheXpert could be used for securely interpreting X-rays with negligible loss of accuracy and in a commercially viable manner. 
This imposes three requirements.
First, there should be automated processes to translate a model, such as DenseNet-121 with several different operators and a very deep network, into an executable for secure inference. 
Second, the translation must result in no deviation between outputs of clear-text and secure inference, ensuring no loss in accuracy or robustness of the model. 
Third, the cost of secure inference as measured by communication and latency should be manageable. 


\begin{figure}
\centering
\begin{tikzpicture}
\begin{axis}[
    height=0.25\textheight,
    xmode=log,
    log ticks with fixed point,
    xmin=75, xmax=3500,
    ymin=0.845,ymax=0.893,
	ytick={0.85,0.86,0.87,0.88,0.89,0.90},
	xtick={100, 200, 400, 800, 1600, 3200},
	every tick label/.append style={font=\tiny},
	ymajorgrids,
	xmajorgrids,
	xminorgrids,
	ylabel={\normalsize Accuracy},
    ylabel style={font=\scriptsize},
    xlabel={\normalsize Communication (GiB)},
    xlabel style={font=\scriptsize}
    ]
    \addplot[
        only marks,
	    mark=*, 
	    paretogreen, 
    ] coordinates {
        (85.62, 0.8677)
	};\label{legend:pareto.green}
	\addplot[		    
	    mark=*, 
	    paretogreen, 
	    point meta=explicit symbolic,
        nodes near coords={{\pgfplotspointmeta}},
        every node near coord/.append style={
            paretogreen,
            font=\tiny, 
            rotate=0, 
            anchor=west
        }
    ] coordinates {
        (85.62, 0.8677) [\ SXW]
        (100.2, 0.8703)  [\ MS]
        (212.407, 0.882) []
	};
	\addplot[		    
	    mark=*, 
	    paretogreen, 
	    point meta=explicit symbolic,
        nodes near coords={{\pgfplotspointmeta}},
        every node near coord/.append style={
            paretogreen,
            font=\tiny, 
            rotate=0, 
            anchor=south
        }
    ] coordinates {
        (212.407, 0.882) [\ R$^{\prime}$F]
        (1701, 0.8884) [\ DDW]
	};
	\addplot[		    
        scatter,
        only marks,
        mark=star,
        paretoblue,
        mark options={scale=1.5},
	    point meta=explicit symbolic,
        nodes near coords={{\pgfplotspointmeta}},
        every node near coord/.append style={
            paretoblue,
            font=\tiny,
            rotate=0, 
            anchor=north
        }
    ] coordinates {
        (2638, 0.8884) [\textbf{DD}]
    };\label{legend:pareto.blue}
    
    
    \node[anchor=north] (dash1start) at (axis cs:200,0.8823){};
    \node[anchor=north] (dash1end) at (axis cs:2900, 0.8823){};
    \node[anchor=north] (dash1text) at (axis cs:800, 0.8817){$\scriptscriptstyle 12.4\times$};
    \draw[dashed,<->](dash1start)--(dash1end);
    
    \node[anchor=north] (dash2start) at (axis cs:80, 0.8677){};
    \node[anchor=north] (dash2end) at (axis cs:2900, 0.8677){};
    \node[anchor=north] (dash2text) at (axis cs:500, 0.8672){$\scriptscriptstyle 30.81\times$};
    \draw[dashed,<->](dash2start)--(dash2end);
    
    \node[anchor=north] (int1start) at (axis cs:2638, 0.8884){};
    \node[anchor=north] (int1end) at (axis cs:2638, 0){};
    \node[anchor=north,scale=0.8] (int1text) at (axis cs:1850, 0.85){\tiny 2638 GiB};
    \draw[densely dotted](int1start)--(int1end);
    
   \node[anchor=north] (int2start) at (axis cs:212.407, 0.8838){};
   \node[anchor=north] (int2end) at (axis cs:212.407, 0){};
   \node[anchor=north,scale=0.8] (int2text) at (axis cs:310, 0.85){\tiny 212.407 GiB};
   \draw[densely dotted](int2start)--(int2end);
    
    \node[anchor=north] (int3start) at (axis cs:85.62, 0.8693){};
    \node[anchor=north] (int3end) at (axis cs:85.62, 0){};
    \node[anchor=north,scale=0.8] (int3text) at (axis cs:120, 0.85){\tiny 85.62 GiB};
    \draw[densely dotted](int3start)--(int3end);
\end{axis}
\end{tikzpicture}
\vspace{-5pt}
\caption{
\textbf{Communication-accuracy Pareto Curve of Proposed models} (marked as 
\ref{legend:pareto.green}): We find a range of models that significantly improve on communication while being accurate relative to SOTA (
marked as \ref{legend:pareto.blue}). Refer to Table \ref{tab:final_results} for the description of model mnemonics used here. 
}
\vspace{-11pt}
\label{fig:pareto}
\end{figure}
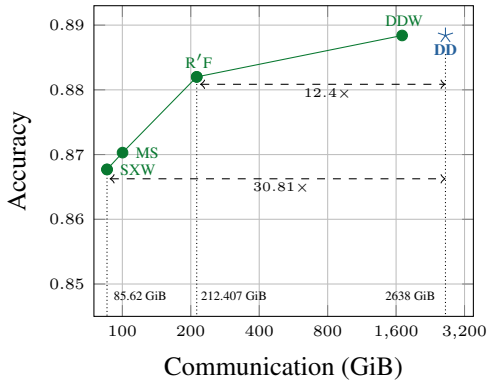

Amongst all existing approaches to secure inference, only the CrypTFlow2  framework satisfies the first two requirements\footnote{We discuss other frameworks in the background section.}. 
The  front-end of CrypTFlow2 can translate a TensorFlow/PyTorch model for secure inference with support for a wide range of common operators.
Also, all secure inference protocols work with fixed-point ML models and CrypTFlow2 provides faithful truncation that ensures bit-wise equivalence between clear-text and secure inference, a highly desirable guarantee on the correctness of secure execution. 
CrypTFlow2 is a state-of-the-art system for secure inference and is the only one to have 
demonstrated accurate and secure inference at this scale (\emph{i.e.} deep models over $320\times 320$ size images) \cite{caring}. Here, CrypTFlow2 provides highly efficient specialized cryptographic protocols.
However, the cost of the resultant secure inference can be still impractically large:
Almost 3TB of communication is required to interpret a single X-ray on DenseNet-121. 
This large overhead remains an outstanding challenge in making secure inference practical.
In this paper, we address the high overheads of secure inference in CrypTFlow2 with three ways to optimize the ML model to be crypto-friendly. \\


\noindent\textbf{1. Efficient Secure Inference of Linear Layers.} We profiled the communication cost of individual layers of DenseNet-121 on CrypTFlow2 and found that linear layers accounted for atleast 96\% of the communication cost (Table~\ref{tab:percentage_comm}). 
This is in contrast to several other studies on efficient secure inference \cite{delphi,cryptonas,safenet,deepreduce} which identified non-linear layers such as ReLU activation as the bottleneck.
CrypTFlow2 provides novel specialized protocols for secure ReLU computation, that significantly reduces their cost, shifting the performance bottleneck to linear layers. 
Since most of the existing techniques on efficient clear-text inference \cite{mobilenet,shufflenet} optimize FLOP-heavy linear layers, we directly adopt them for secure inference. 
In particular, we adopt optimized network backbones, factorized depth-wise separable convolution, and Shuffle operators. 
We show that communication cost can be significantly reduced (1.5$\times$ to over 8$\times$) by reducing FLOPs, while achieving high accuracy. \\
\textbf{2. A custom Crypto-friendly X-Operator.}
Going beyond existing techniques,
we design a new custom crypto-friendly X-operator as a drop-in replacement for the {\em bottleneck cells} \cite{resnet}. 
The X-operator is composed of grouped point-wise convolution, followed by an addition on the outputs of ``parallel'' shuffle and depthwise convolution, followed by another grouped point-wise convolution. 
For the MobileNetV3-Large backbone, 
the X-operator reduces the communication cost by over 90\% while incurring only a 0.011 drop in AUC (Table~\ref{tab:final_results}). \\
\textbf{3. Winograd Algorithm for Convolutions.}
While the above two optimizations change the model by introducing efficient operators, our third contribution is a compiler technique whereby we rewrite convolutions to use the fast Winograd algorithm \cite{winograd}.
This reduces the number of multiplications, which are expensive in secure inference, while increasing number of additions which do not incur any communication. 
We add support for the Winograd algorithm as part of the Athos~\cite{cryptflow} compiler in CrypTFlow2 to automatically optimize any linear layer.
As an example, for the DenseNet-121 network, the Winograd algorithm reduces communication cost by 55\% without any loss in accuracy.
We show similar results across other network backbones as well. 

\subsubsection{Our Results.}
In combination, these optimizations result in a range of models\footnote{We consider only non-ensemble models in this work. We expect the benefits to translate for ensembles.} panning a Pareto curve of accuracy and inference cost on the CheXpert dataset, as shown in Figure~\ref{fig:pareto}. 
For instance, we identify a model with about 12.4$\times$ lower communication and a negligible drop in AUC of 0.006, and another with over 30$\times$ lower communication (bringing the communication  below 100 GB and latency below 90s) with a drop in AUC of 0.02. 
In contrast, the architecture optimizations proposed in prior work~\cite{delphi,cryptonas,safenet,deepreduce} can reduce communication cost by only $1\%$ to $4\%$ for CheXpert.
We also note that the problems of efficient clear-text inference and efficient secure inference are distinct as the optimizations we consider don't offer commensurate benefits for clear-text inference. 
For instance, our model that lowers secure inference cost by 30$\times$, improves clear-text latency on an Intel Xeon CPU by only 7$\times$ and a V100 GPU by only 4$\times$.  
Finally, our significant cost reductions for secure inference demonstrate that crypto-friendly ML models can be designed to make secure inference practical for realistic models.

\subsubsection{Other Related Works.}
There are two primary approaches to improve the performance of secure inference on a given task: changing cryptography or changing the model. 
These two approaches are complementary.
In the latter category, \cite{delphi,cryptonas,safenet,deepreduce} reduce the number of ReLUs.
Cryptonets~\cite{cryptonets} and ngraph-HE~\cite{nhe} replace ReLUs entirely by quadratics.
CoiNN~\cite{coinn} and SiRNN~\cite{sirnn} use varying bitwidths to balance accuracy and efficiency. 
Reducing bitwidths is complementary to our work and improves the efficiency of secure evaluation of both linear and non-linear layers.

\section{Background}
\label{sec:background}
Here, we provide the necessary background on secure 2-party computation (2PC) and the cost profile of different operations with the state-of-the-art protocols in the CrypTFlow2 framework  that guide our operator choices in this paper for secure inference. 

\subsubsection{2PC and Secure Inference.}
2PC \cite{yao,gmw} allows two parties $P_0$ and $P_1$ holding private inputs $x$ and $y$, respectively, to {\em securely} compute $f(x,y)$ for any public function $f$. It provides a formal security guarantee that a party does not learn anything about the other party's private input beyond what can be deduced from the function output.  It is easy to see that secure inference can be solved using 2PC between the model owner, say $P_0$, holding the model weights $w$ and the client, say $P_1$, holding the data point $x$ who want to compute $M(w,x)$, where $M$ is the public model architecture. 

All works on 2-party secure inference use cryptographic primitive called {\em additive secret sharing} that allows to split private data (\emph{secret}) into 2-parts called {\em secret shares} such that a single share completely hides the secret and the secret shares can be added to reveal the secret. Parties start by secret sharing their inputs, and then for each layer of the network, parties start with secret shares of the input to that layer and run a protocol to securely compute the secret shares of the output of that layer. All works on secure inference provide such protocols for both linear and non-linear layers. These protocols require communicating bits that are indistinguishable from random, which makes it secure.


\subsubsection{Cost characteristics of secure inference protocols.}
2PC protocols are interactive \emph{i.e.,} both the parties exchange cryptographic messages over multiple rounds of interaction. And for secure inference of large models, the cost of the secure inference is governed by the communication of the protocol, \emph{i.e.}, the total amount of data being exchanged between the two parties. It is important to note that cost of different operators in 2PC is quite different from corresponding cost over the clear-text (that is, when the inputs are known in the clear for computation). For instance, while  additions and multiplications cost roughly the same over clear-text, in 2PC additions are free and multiplications are expensive. This is because the secret sharing scheme is additive in nature and \emph{no} interaction is needed for adding two secret values. 
Similarly, any operation that permutes the data in a tensor requires no interaction since both parties can individually permute their shares of the tensor, e.g., shuffle operator. 

On the other hand, non-linear operations such as ReLU and Maxpool require interaction. Works prior to CrypTFlow2 such as MiniONN~\cite{minionn}, Gazelle~\cite{gazelle}, Delphi \cite{delphi} and others used garbled circuits~\cite{yao} for these functions that were communication intensive. 
CrypTFlow2 provides novel specialized 2PC protocols for non-linear operations that reduce their cost by more than $10\times$. 

\subsubsection{ML to 2PC.} 
Directly working with 2PC protocols requires expertise in cryptography. Hence, in recent years, several frameworks have been developed that automatically generate secure inference implementations from the clear-text ML code in various threat models~\cite{cryptflow,quantizednn,tfe,pysyft,crypten}.
The users train ML models in Keras, PyTorch, or Tensorflow and use these frameworks to get secure inference implementations at the push of a button. 
Among these, only CrypTFlow2 generates 2PC protocols; the other frameworks require additional trust assumptions. 
Furthermore, CrypTFlow2 guarantees correctness, \emph{i.e.}, the the clear-text execution is bitwise equivalent to the generated 2PC implementation. This guarantee is critical for the feasibility of our evaluation. 
Since 2PC protocols are expensive, it is intractable to run 2PC implementation for evaluating accuracy on various backbones and optimized networks. 
Armed with this correctness guarantees, we use the clear-text implementations to measure accuracy soundly.

\section{Characterising Secure Inference of DNNs} 

Deep models for computer vision, such as DenseNet-121, ResNet-50 \cite{resnet}, MobileNetV3-Large \cite{mobilenet}, and ShuffleNetV2 \cite{shufflenet} contain a large number of layers and various operators, which may be classified as either linear or non-linear.
Linear layers consist of operations such as convolutions, matrix multiplications, batch normalization and use additions/multiplications.
Non-linear layers consist of activation functions such as ReLU, pooling layers such as max-pool, and classifier layers such as argmax.
We profile the communication cost of secure inference for these common deep models, and identify linear layers (particularly convolutions) to be the bottleneck.

We profile the four models with an input image size of $320\times 320$ using CrypTFlow2, instrumented to record communication cost at an operator level.
We report the split of communication cost for linear and non-linear layers in Table \ref{tab:percentage_comm}. 
We find that, across the four models, at least 96\% of the communication is attributed to linear layers%
\footnote{
Although these costs are obtained using oblivious transfer based 2PC protocol in CrypTFlow2, we believe that our techniques should give similar benefits when executed with SOTA HE-based protocols~\cite{cheetah} where bottlenecks are similar.
}.
This dominance of the cost of linear layers in secure inference differs from prior work which found non-linear layers to be more expensive~\cite{delphi,cryptonas,deepreduce}. 
The reasons for this difference are two-fold: 
1) Prior work primarily focused on models for small images such as in MNIST and CIFAR datasets (up to 32x32) wherein the linear layers are less computationally expensive. 
In contrast, we work with realistic images such as X-ray images of size 320x320.
2) Prior work used secure inference protocols with Garbled Circuits (GC) \cite{yao} which are communication intensive for non-linear layers \cite{delphi}.
In contrast, the current SOTA secure inference methods such as CrypTFlow2 provide specialized protocols for non-linear layers that are over $10\times$ cheaper than GC.

\begin{table}[!t]
\centering
\scalebox{0.7}
{
\begin{tabular}{l|c|c|c|c}
\toprule 
Network & \multicolumn{3}{|c|}{\% of linear} & \% of non-linear \\ 
\toprule 
& \% of convolution & \% of other linear & \% all & \\
\toprule 
DenseNet-121 & 97.11\% &  1.58\% & 98.69\% &  1.31\% \\ 
ResNet-50 & 96.67\% & 1.89\% & 98.56\% & 1.44\% \\ 
MobileNet-V3 & 90.56\% & 5.72\% & 96.28\% & 3.72\% \\ 
ShuffleNet-V2 & 92.82\% & 4.10\% & 96.92\% & 3.08\% \\
\toprule
\end{tabular}
}
\caption{\textbf{Split of communication across linear and non-linear layers.} Across DNNs, linear layers dominate communication. Amongst them, convolution layers dominate.}
\label{tab:percentage_comm}
\vspace{-12pt}
\end{table}

Within linear layers, we examine the communication cost between convolution layers and other layers (including batch normalization). We find that convolutions account for at least $90\%$ of the communication cost in all our computer vision models (Table~\ref{tab:percentage_comm}). 
Again, this is in contrast to equivalent numbers for clear-text inference: In MobileNetV3-Large for instance, convolutions account for about 70\% of the computational cycles when running on a CPU. 

The above profiling results indicate opportunities for making secure inference efficient. 
In contrast to prior work with other protocols, CrypTFlow2's communication cost is localized within linear layers. 
And in contrast to clear-text inference, within linear layers, convolutions are most expensive in CryptFlow2. 
Furthermore, tensor additions are local operations and thus communication-free, while multiplications are expensive and require 1.2KB of communication per operation. 
Thus, \textbf{reducing the number of multiplication operations}, denoted $\mult$, emerges as an effective strategy for efficient secure inference.

\section{Efficient Alternatives to Convolution}
Designing deep models for efficient clear-text inference has been a widely studied problem \cite{efficientdl}. 
Such attempts have to consider several aspects such as reducing floating-point operations (FLOPs), optimizing for a given architecture, exploiting locality in memory access patterns, and so on. 
However, the above profiling of communication cost for secure inference suggests a simple strategy of reducing $\mult$s, which are situated primarily in convolution operations. 
Consequently we consider two existing approaches to optimize convolution operations: (a) factorized convolution, and (b) multiplication-free shuffle operation. We also demonstrate out-sized reductions in cost for secure inference compared to clear-text with these optimizations.



\subsubsection{Factorized Convolution.} 

\begin{figure}
    \centering
    \includegraphics[width=0.4\textwidth]{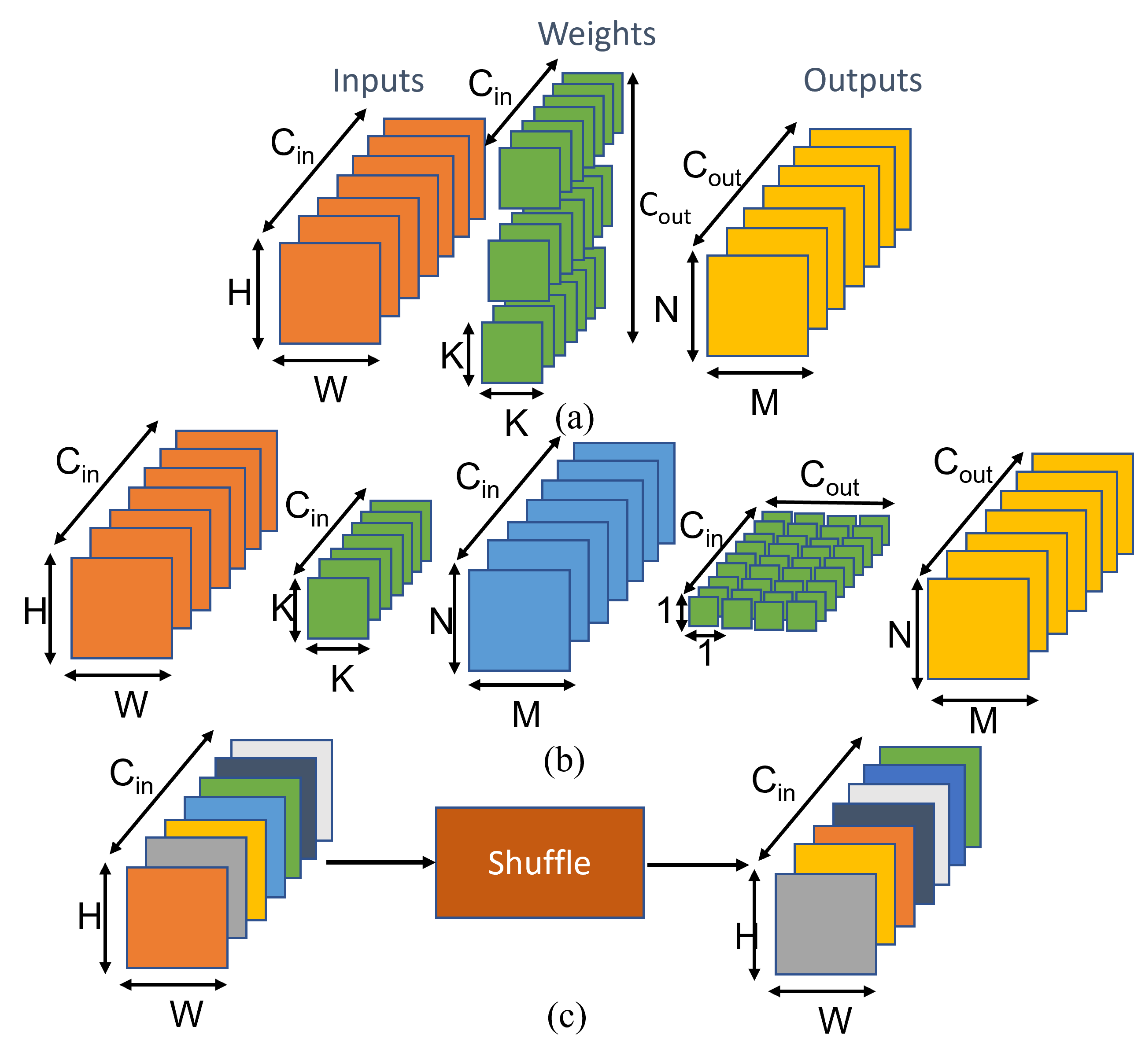}
    \vspace{-5pt}
    \caption{\textbf{Operators used in clear-text.} (a) Dense Conv, (b) Factorized depthwise-separable Conv, and (c) Shuffle}
    \vspace{-12pt}
    \label{fig:sp_dw_conv}
\end{figure}

Convolutions form a key component in models for computer vision, and are designed to aggregate information across spatial (usually two: x- and y-axes) and depth or channel (usually many) axes. 
Like tensor multiplication, convolution can also be factorized, along spatial axes \cite{squeezenext}, depth axis \cite{xception,mobilenet}, or both \cite{scaffoldingdl, fuseconv}.
Indeed, most efficient models (that are deployed in low-resource devices such as mobile phones) use depthwise separable convolutions (DS-Convolution), as shown in Figure~\ref{fig:sp_dw_conv}(b).
The aggregation of information along spatial and depth axes are separated out: 
Independent 2-D convolutions on each input depth are followed by pointwise or $1\times 1$ convolutions.

Depthwise factorization can significantly reduce $\mult$.
For instance, a standard convolution with a filter of size $\cout\times \cin\times K\times K$ on an input of size $\cin\times H\times W$ and output of size $\cout\times N\times M$ requires $NM\cout K^2\cin$ $\mult$.
If this was converted into a DS-Convolution, it would require $NM\cin (K^2 + \cout)$ $\mult$, a reduction by a factor of $\frac{K^2\cout}{K^2+\cout}$.
As an example, for a filter with a spatial size of $K = 3$ and input features of size $\cout = 32$, moving from dense to DS-convolution can reduce $\mult$ by over $7\times$.

Do the reduction in $\mult$ with factorized convolution translate to lower communication costs for secure inference? 
We empirically evaluate this with a convolution operator with $K = 3, \cin = 3, \cout = 64$, and various input sizes. 
We compute the communication with CrypTFlow2 for both dense and DS-convolution while varying the image size from $16\times 16$ to $512\times 512$.
The results, shown in Figure \ref{fig:op_scaling}, indicate a consistent 5$\times$ reduction in communication cost for DS-convolution across image sizes.
However, this reduction in communication cost needs to be traded-off with any drop in accuracy.  
We evaluate this for models on the CheXpert dataset and show that the accuracy loss is negligible.

\subsubsection{Shuffle Operation.} 
In channel-wise Shuffle \cite{shufflenet}, 
channels of a feature map are randomly permuted, enabling information flow across channels (see Figure \ref{fig:sp_dw_conv} (c)). 
When implemented for clear-text inference, permuting channels has memory and thus latency overheads.
However, for 2PC, Shuffle does not add $\mult$ and is a local operation that is communication-free.
Thus, for secure inference, we can reduce the number of $\mult$ by sharing the compute available between convolution and shuffle. 
Recall that DS-convolution has $\cin$ independent 2-D depthwise convolutions followed by channel aggregating $1\times 1$ pointwise convolution. 
When using depthwise convolutions in composition with Shuffle, $\cin/2$ channels are randomly permuted while the other $\cin/2$ channels have independent 2-D convolutions reducing $\mult$ by half with respect to DS-convolution. 

\begin{figure}
\centering
\hspace{-25pt}
\begin{tikzpicture}
\begin{axis}[
    height=0.24\textheight,
    xmode=log,
    ymode=log,
    log ticks with fixed point,
    xmin=12, xmax=656,
    ymin=0.01,ymax=500,
	ytick={0.05, 0.1, 0.5, 1, 5, 10, 50, 100, 500},
	xtick={16, 32, 64, 128, 256, 512},
	every tick label/.append style={font=\tiny},
	ymajorgrids,
	xmajorgrids,
	xminorgrids,
	ylabel={\normalsize Communication (GiB)},
    ylabel style={font=\scriptsize},
    xlabel={\normalsize Image size},
    xlabel style={font=\scriptsize},
    legend pos = south east,
    legend style={nodes={scale=0.75}}
    ]
	\addplot[		    
	    mark=o, 
	    myblue
    ] coordinates {
        (16, 0.213982)
        (32, 0.844094)
        (64, 3.35898)
        (128, 13.4265)
        (256, 53.6953)
        (512, 214.77)
	};
	\addplot[
	    mark=x,
	    myred
	] coordinates {
	    (16, 0.0426492999999999)
        (32, 0.16905)
        (64, 0.672624)
        (128, 2.68698)
        (256, 10.7446)
        (512, 42.9785)
	};
	\addplot[
	    mark=*,
	    mygreen
	] coordinates {
	(16, 0.023694055555555496)
    (32, 0.09391666666666666)
    (64, 0.37368)
    (128,1.4927666666666668)
    (256,5.969222222222222)
    (512, 23.876944444444444)
	};
\addlegendentry{{\small Dense Conv}}
\addlegendentry{{\small Factorized Conv}}
\addlegendentry{{\small Shuffle}}
\end{axis}
\end{tikzpicture}
\vspace{-5pt}
\caption{\textbf{Scaling of efficient clear-text operators.} We observe consistent communication benefits with factorized convolution and shuffle for increasing image sizes
}
\vspace{-13pt}
\label{fig:op_scaling}
\end{figure}
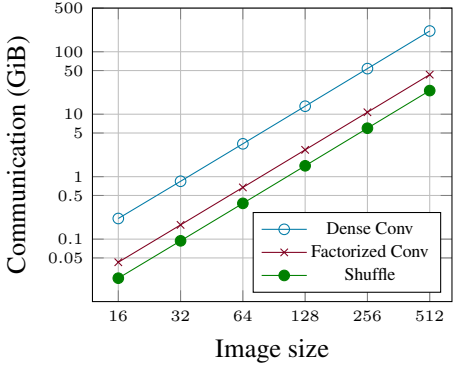

Just as with DS-convolution, we empirically compute the communication cost of the Shuffle operator for various input image sizes. 
We see that moving from DS-Convolution to Shuffle reduces the communication cost by $1.8\times$, a reduction that is consistent across image sizes. (see Figure \ref{fig:op_scaling})

\subsubsection{Gains in Clear-text vs Secure Inference.}
We showed that DS-convolution reduced communication costs by 5$\times$ relative to dense convolution, and Shuffle reduced it by a further $1.8\times$ for an aggregate saving of $9\times$.
For clear-text inference, DS-convolution reduces the average execution time on a CPU only by $1.4\times$ and GPU by only $2\times$ relative to dense convolution, with shuffle not improving the execution time any further.
These out-sized gains in secure inference suggest a direct and sensitive dependence on $\mult$ while efficient clear-text inference depends on multiple factors including the architecture on which it is deployed. 

\section{A Custom Efficient Operator for 2PC}
\label{sec:shuffled_group}
\noindent
In the previous section, we showed that reducing $\mult$ is an effective strategy to reduce communication cost for secure inference, and that existing efficient operators for clear-text inference provide out-sized gains.
Next, we design a custom operator specifically suitable for secure inference.




\subsubsection{Factorized Point-Wise Convolution.}

\begin{figure}
    \hspace{-20pt}
    \includegraphics[width=0.54\textwidth]{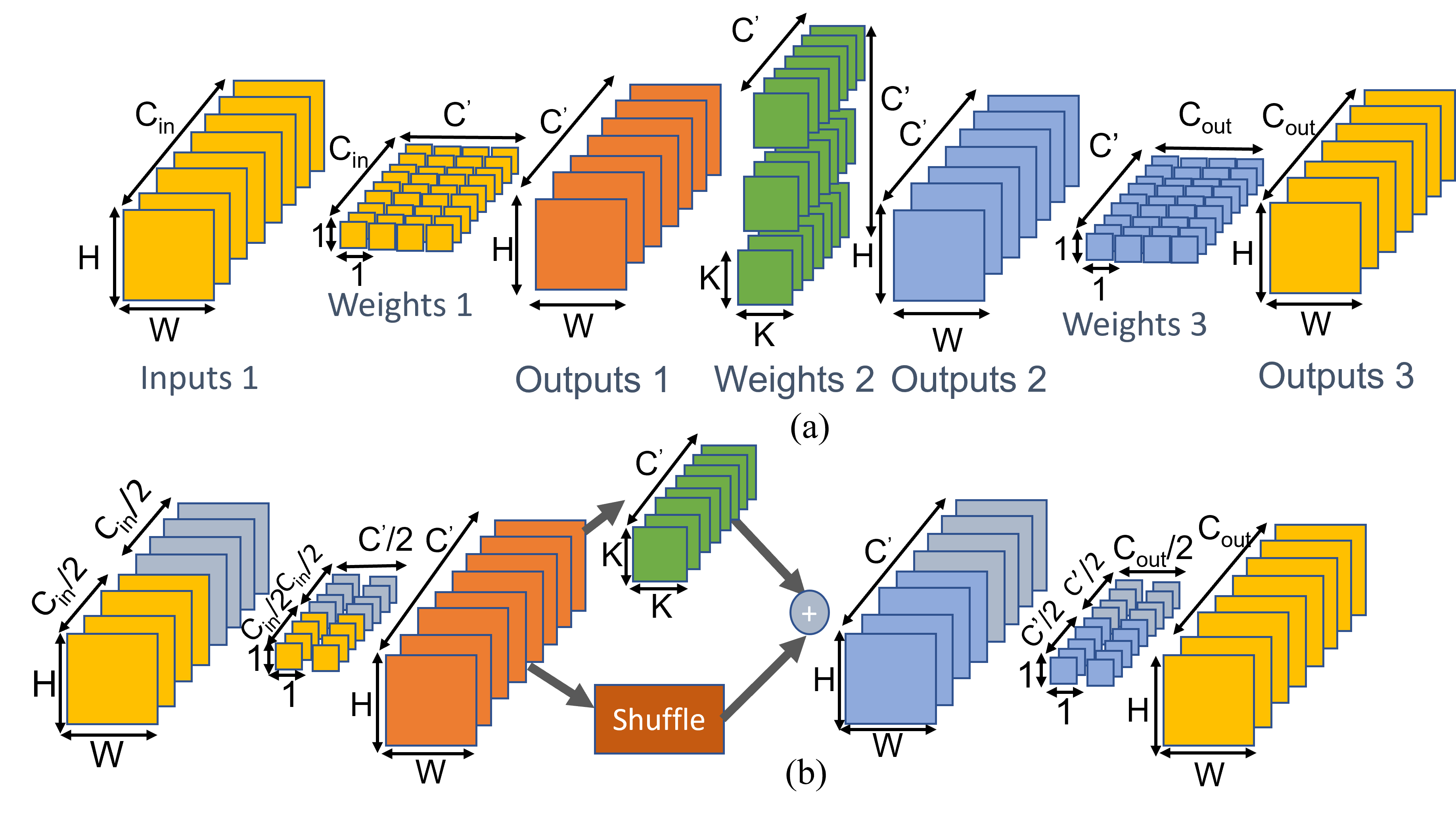}
    \vspace{-12pt}
    \caption{(a) Bottleneck cell, (b) X-operator. The proposed operator optimises both pointwise and dense convolutions with factorisation and parallel composition.} 
    \vspace{-10pt}
    \label{fig:gsb}
\end{figure}

Current deep models for computer vision use a cell-based design, i.e., a basic building-block is identified and then replicated with different dimensions in various layers to build a deep model.
For instance, bottleneck cells have been commonly used since they were shown to be effective initially for ResNet-50 \cite{resnet}. 
A bottleneck cell contains a $K\times K$ dense convolution  sandwiched between two point-wise convolutions (Figure~\ref{fig:gsb} (a)). 
As discussed earlier, efforts have been made to make dense convolution more efficient by using factorization or Shuffle operators. 
Point-wise convolutions are expressible as matrix multiplications and have not received much attention for efficiency as they do not dominate clear-text inference costs.

To understand the corresponding costs for secure inference, we profiled a bottleneck cell (see Figure \ref{fig:gsb} (a)) with dense, DS-convolution and  Shuffle  for a feature map of size $3\times 320\times 320$, a dense filter of size $3\times 3$ and $\cin=3, \cout=64$ with intermediate channels $C^{\prime}=48$. 
The fraction of the communication cost attributed to point-wise convolution for the three cases are $13.4\%$, $89.3\%$, and $95\%$. 
Thus in these optimized operators, the performance bottleneck 
has shifted to the point-wise convolution. 

To address this, we propose to novel factorized point-wise convolution for 2PC. Consider a point-wise convolution of an input of size $\cin \times H \times W$ with a filter of size $\cout \times \cin \times 1 \times 1$. 
This entails $HW$ tensor dot-products of tensors of size $\cin$ for every $\cout$ filter, a total of $\cout\cin HW$ $\mult$.
We propose to factorize this into two groups, i.e., divide the input map into two tensors with $\cin/2$ channels each and perform two independent point-wise convolution with two filters of size $\cin/2$, followed by a tensor composition. With this, we obtain 2 outputs per $\cin$ multiplications. Hence, to maintain the same output feature map size, we reduce the number of filters to $\cout/2$, effectively cutting the overall $\mult$ by half to $\cout \cin HW/2$.
%

\subsubsection{``Parallel'' composition of Shuffle and Depth-wise convolution.}
As discussed earlier, in secure inference both shuffle operations and additions of tensors are communication-free. 
Thus, they can be used to express information flow across channels and aggregate information respectively without any cost.
Recall that depthwise convolutions perform independent 2-D convolutions on each channel. 
We propose to augment depthwise convolution by adding a ``parallel'' path where shuffle operation is also performed on \textit{all} channels enabling information flow across channels. 
Then the results of the shuffle and convolutions paths are aggregated with a tensor addition, which does not incur communication.
Such parallel composition is expected to double the latency cost of clear-text inference on CPUs, but is uniquely efficient for 2PC.

\subsubsection{X-operator.} We combine the above two propositions into a single operator that we call the X-operator, as visualized in Figure~\ref{fig:gsb}.
It comprises of a parallel shuffle and depthwise convolution, that is sandwiched between two point-wise convolutions that are factorized into two groups.
The X-operator can be a drop-in replacement for a bottleneck cell.
For instance, in layer 1 of ResNet-50, the bottleneck cell uses dense convolution and has 1.46B $\mult$. 
Replacing the dense convolution within the bottleneck with DS-Convolution and Shuffle, the $\mult$ is reduced to 0.53B ($2.75\times$).
Instead, replacing the bottleneck layer with the X-operator reduces $\mult$ to 0.27B ($5.3\times$).


\subsubsection{X-operator for Secure Inference.}
We empirically compute the communication cost of X-operator in secure inference with parameters $\cin=16$, $\cout=64$ with $K=3$. The input and output channels for the sandwiched bottleneck $C^\prime = 48$.
We observe that X-operator gives a reduction in communication cost of $1.5\times$ relative to bottleneck cell with half DS-convolution and half Shuffle, and $7.59\times$ relative to bottleneck cell with dense convolution, consistently across image sizes. 
This establishes the effectiveness of designing custom operators that are crypto-friendly. 

Here, we note the value of an automated compiler flow.  
We designed a custom-operator network that can be trained on a large dataset, and were able to compile it into a secure executable through Athos. 
This automated flow is crucial to accelerate the discovery of potentially many other crypto-friendly operators.

\section{Winograd Algorithm for Convolutions}
\label{sec:wino}
\noindent 
In the previous two sections, we proposed ways to reduce the cost of secure inference with existing and custom operators.
The adoption of these operators is conditioned on the accuracy of models using them.
In this section, we propose an alternative algorithm to implement convolution operator while retaining functional equivalence.

Given the dominance of convolution operation in deep models, several methods exist to improve its performance.
In particular, the Winograd algorithm \cite{winograd} implements convolution with fewer multiplications at the cost of increased number of additions and storage of intermediate outputs. 
In clear-text inference, Winograd algorithm has not been shown to reduce inference cost significantly, due to the relatively similar costs of addition and multiplication and also the cost of additional memory accesses \cite{winogradhandbook, winograd}. 
However, for secure inference Winograd algorithm can effectively reduce communication cost given that it reduces $\mult$.


\subsubsection{Winograd Algorithm in Secure Inference.}
Consider the convolution of an image $I$ of size $N \times N$ with a filter $F$ of size $K \times K$  to obtain an output $O$ of size $M \times M$. \footnote{Note that the relation $N=M+K-1$ holds}
In dense convolution this requires $M^2K^2$ $\mult$.
With the Winograd algorithm, the same operation can be computed with $(M + K - 1)^2$ $\mult$. 
As an example, for output image of size $5\times 5$ and filter of size $3 \times 3$, the reduction in $\mult$ is about $4.5\times$.
Formally, the algorithm computes: 
    $$O = A^T[(B^TIB)\odot(GFG^T)]A$$  
where $(A, B, G)$ are constant and public tensors dependent only on the image/filter sizes and can be pre-computed.
$\odot$ is element-wise multiplication.
For secure inference, the terms $B^TIB$ and $GFG^T$ can be computed locally due to linearity of secret sharing. 
Evaluating $\odot$ requires $(M+K-1)^2$ element-wise multiplications, which incurs 2PC communication cost. 
The final $A^T[\cdot]A$ can also be computed locally without any communication cost. 

\subsubsection{Tiling.}

Using the Winograd algorithm for convolving a large image in a single shot causes numerical errors because of the presence of irreducible fractions in the $(A, B, G)$ matrices mentioned in the main text. The $\mult$ for convolving an $M\times M$ image with a $K\times K$ using Winograd without tiling
\begin{align*}
    \beta_1(M, K) = (M+K-1)^2
\end{align*}
To convolve a large image we split it into multiple equal sized tiles and perform Winograd convolution for each input tile. 
Subsequently, the outputs tiles are then pieced together to obtain the complete image. With tiles of size $n\times n$, the number of tiles required to cover the image along an axis is
\begin{align*}
    T=\left\lceil\frac{M-K+1}{n-K+1}\right\rceil
\end{align*}
Convolving each tile invokes $n^2$ multiplications. Thus, $\mult$ for the tiled Winograd algorithm is
\begin{align*}
    \beta_2(M, K; n) = T^2n^2
\end{align*}
It is straightforward to show that for all image sizes $M\geq n$, the multiplications invoked without tiling is lesser than with tiling \emph{i.e.,}
\begin{align*}
    \beta_1(M, K) \leq \beta_2(M, K; n)
\end{align*}
For instance, for a $320\times 320$ image $3 \times 3$ filter, $\mult$ for standard convolution is 910,116 and non-tiled Winograd algorithm is 102,400. On the other hand, the tiled Winograd algorithm with tile size $6\times 6$ would use 230,400 multiplications.

\subsubsection{Details of Tiling.}
Images with arbitrary size $M\times M$ need not be divisible neatly into tiles of size $n\times n$. The right/bottom edges of the image may need to be covered by tiles of a smaller size $n'\times n'$. Here $r\leq n' \leq n-1$. For ease of implementation, we pad these edges of the image with 0 so that a \emph{single} tile size $n\times n$ can evenly cover the image. However, padding incurs an increase in the number of multiplications by a factor $\gamma$, which we estimate
\begin{align*}
    \gamma &= \frac{\text{\#}\mathsf{multiplications\ with\ padding}}{\text{\#}\mathsf{multiplications\ without\ padding}} \nonumber \\
    &= \frac{T^2n^2}{[(T-1)n + n']^2} = \left[\frac{1}{1-(\frac{1-n'/n}{T})}\right]^2
\end{align*}
This factor is maximized when $n'=K$. Under a first-order approximation, we have
\begin{align*}
    \gamma \approx 1 + 2\left(\frac{1-K/n}{T}\right)
\end{align*}
Convolving large images entails $T>>1$ and hence the overhead factor $\gamma$ due to padding is negligible.

\subsubsection{Compiler support.}

In the Athos compiler front-end, we enable an ML developer to specify whether a convolution operation is to be implemented as a Winograd algorithm. This automated feature enables a functionally equivalent implementation while reducing the 2PC communication cost. 
%
\subsubsection{Limitations.} 

Winograd transformation can only be applied to convolutions with a square $K\times K$ filter with $K>1$. Therefore if the network is dominated by pointwise convolutions - the gains diminish. Moreover, the nature of the Winograd transformation is only applicable to convolutions with stride $s=1$. Making the Winograd transformation work for stride $s>1$ is an open research problem with multiple solutions having been proposed \cite{multi.wino1, multi.wino2, multi.wino3} in the recent past. Using Winograd convolution for $r=7$ causes a lack of numerical precision, and hence we avoid it. 
Due to the above technicalities, while this transformation may not always be applicable, it still gives promising gains when applied to networks with large dense convolution layers.

\section{Experimental Evaluation} 
\label{sec:evaluation} 
\noindent

\noindent In this section, we share the methodology and results from the experiments evaluating our proposed optimizations.
\subsection{Methodology} 
\subsubsection{Models.}
We evaluate the optimizations with five different models that we have discussed in the paper thus far: DenseNet-121, ResNet-50, ResNet-18, MobileNetV3-Large, and ShuffleNetV2. 
We consider different variants of these models, starting with a baseline variant where all convolution operations are dense convolutions. 
Then, we progressively use the optimized operators: DS-Convolution, \shuffle, and X-operator.
Most of these models (except ResNet-18) use \textit{bottleneck cell} as the basic building block, \emph{i.e.} a dense convolution sandwiched between two pointwise convolutions. As we \textit{factorize} the dense convolution into DS-Convolution, one can observe two back to back $1\times 1$ pointwise convolutions, one arising from the factorization while the other existing within \textit{bottleneck cell}. We combine these two repeating pointwise convolutions for parameter efficiency.
For the most efficient model family, ShuffleNetV2, we optimize each variant with the Winograd algorithm.
For all other models, Winograd is applied only on the dense variant where it is found to be most effective.


\subsubsection{Training setup.}
We use PyTorch \cite{pytorch} to train the models on the CheXpert~\cite{chexpert} challenge training dataset consisting of 224,316 chest radiographs from 65,240 patients and evaluate them on the validation dataset from 200 patients involving 5 output classes : (a) Atelectasis, (b) Cardiomegaly, (c) Consolidation, (d) Edema, and (e) Pleural Effusion.
The training labels of the dataset can be 0, 1, or uncertain (u). 
In practice, the u labels are converted to a 1 (U-Ones), 0 (U-Zeros), or ignored (U-Ignore). 
We convert the uncertain labels to 1 and utilise the U-Ones model throughout for training. 
We apply standard pre-processing for CheXpert with a center-crop of $320\times320$ and normalise the pixels with a mean of $0.533$ and standard deviation $0.0349$.
We train model variants to maximise the area under the receiver operating curve (AUC) and report that as the accuracy metric. We use SGD as the optimizer with an initial learning rate of $0.001$.
We use FP32 precision and train the models for 10 epochs on a V100 cluster with a batch size of 16.

\subsubsection{Secure inference.} Models trained in PyTorch are exported into  ONNX \cite{onnx} which is then processed by the Athos toolchain to output fixed-point models with bit-width 60 and scale 23.
These parameters ensure that the accuracy of fixed-point models matches that of PyTorch. The clear-text fixed-point models are linked with the secure protocols 
in CrypTFlow2.
We measure latency on a LAN setup where two VMs are connected by a network with 16Gbps bandwidth and 0.7 ms round-trip time. The two VMs have the following specification.
\begin{table}[!ht]
\centering
\begin{tabular}{|c|c|}
\hline
\textbf{Processor}   & AMD EPYC 7763 64-Core \\ \hline
\textbf{Clock Speed} & 2.44 Ghz              \\ \hline
\textbf{RAM}         & 128 GB                \\ \hline
\textbf{Bandwidth}   & 16 Gbps               \\ \hline
\textbf{Ping Time}   & 0.7 ms                \\ \hline
\end{tabular}
\end{table}


\begin{table}[!t]
\hspace{-7pt}
\scalebox{0.6}
{
\begin{tabular}{lcccrrc}
\toprule 
Backbone & Operator & Winograd & Mnemonic & Communication & Latency & Average AuC \\
& & & & Reduction & Reduction & on CheXpert \\
\toprule
             & Dense &  & DD & $1\times$ & $1\times$ & 0.888 \\
             & Dense & \checkmark & DDW  & $1.55\times$ & $1.06\times$ & 0.888 \\
DenseNet-121 & Factorized &  & DF & $1.56\times$ & $1.06\times$ & 0.885 \\
             & Shuffle &  & DS & $1.57\times$ & $1.06\times$ & 0.886 \\
             & X-op & & DX & $2.37\times$ & $1.33\times$ & 0.880 \\
\toprule
             & Dense & & RD & $0.71\times$ & $0.76\times$ & 0.884 \\
             & Dense & \checkmark & RDW & $0.96\times$ & $0.87\times$ & 0.882 \\ 
ResNet-50    & Factorized & & RF & $1.18\times$ & $1.08\times$ & 0.883 \\
             & Shuffle & & RS & $1.18\times$ & $1.17\times$ & 0.882 \\  
             & X-op  &  & RX & $2.10\times$ & $1.43\times$ & 0.876 \\
\toprule 
            & Dense & & R$^{\prime}$D & $1.61\times$ & $1.90\times$ & 0.884 \\
             & Dense & \checkmark & R$^{\prime}$DW & $3.61\times$ & $3.31\times$ & 0.884 \\
ResNet-18    & Factorized & & R$^{\prime}$F & $12.41\times$ & $7.27\times$ & 0.883 \\
             & Shuffle & & R$^{\prime}$S & $12.83\times$ & $8.25\times$ & 0.882 \\
             & X-op & & R$^{\prime}$X & $19.10\times$ & $8.77\times$ & 0.866 \\
\toprule 
                  & Dense & &  MD & $2.18\times$ & $2.33\times$ & 0.882 \\
                  & Dense & \checkmark &  MDW & $8.30\times$ & $2.91\times$ & 0.882 \\
MobileNetV3 & Factorized & &  MF & $12.93\times$ & $9.11\times$ & 0.873  \\
Large                  & Shuffle & &  MS & $18.30\times$ & $9.11\times$ & 0.870 \\
                  & X-op & &  MX & $19.39\times$ & $9.91\times$ & 0.871 \\
\toprule 
           & Dense & & SD & $5.50\times$ & $2.11\times$ & 0.879 \\
           & Dense & \checkmark & SDW  & $9.07\times$ & $7.30\times$ & 0.879 \\
           & Factorized & & SF  & $19.39\times$ & $12.84\times$ & 0.872 \\
ShuffleNetV2 & Factorized & \checkmark  & SFW & $19.66\times$ & $10.30\times$ & 0.872  \\
           & Shuffle & &  SS & $19.67\times$ & $12.84\times$ & 0.872 \\
           & Shuffle & \checkmark & SSW  & $19.67\times$ & $12.84\times$ & 0.875 \\
           & X-op & &  SX  & $30.15\times$ & $15.35\times$ & 0.868 \\
           & X-op & \checkmark & SXW & $30.80\times$ & $17.09\times$ & 0.868 \\
\toprule 
\end{tabular}
}
\caption{Evaluating the benefits of models supporting different operators. The gains are relative to DenseNet-121 (DD) with 2.63TB of communication and 1524.9s latency.}
\vspace{-14pt}
\label{tab:final_results}
\end{table}

\subsection{Results}
\vspace{-1.8pt} 
In Table \ref{tab:final_results} we report the accuracy on CheXpert and relative gains in communication and latency costs with respect to DenseNet-121 for inference of a single input image on each of the model variants. 
We make the following observations: \\
\textbf{Choice of model.} For the Dense variants, the choice of the model itself affects the cost - ResNet-50 incurs more communication than DenseNet-121 while MobileNetV3-Large leads to a $2.18\times$ reduction for a reduction in AUC by 0.006.\\
\textbf{Factorized DS-convolution.} Factorizing dense convolution to DS-convolutions reduces communication cost by an average of $4.07\times$ across models.
For instance, ResNet18 has $7.7\times$ lesser communication (R$^\prime$F vs R$^\prime$D) with very negligible loss in AUC.
Gains are however not uniform across models --- DenseNet-121 and ResNet-50 improve modestly.\\
\textbf{Shuffle Operator.} Using the Shuffle operator which optimizes DS-convolution further by using shuffle for half the channels leads to a marginal reduction in communication cost, by an average of $1.1\times$ w.r.t. to the corresponding model with factorized convolution.
Specifically, $MS$ has about $1.41\times$ lower communication cost than $MF$ with a drop in AUC of 0.003. 
Models which had lower gains with factorized convolution also have lower gains with Shuffle. \\
\textbf{X-Operator.} Using the X-operator provides the largest gains for all models by an average of $8.88\times$ w.r.t. to corresponding models with dense convolutions. 
Specifically, $SX$ has a $30\times$ lower communication cost than the baseline DenseNet-121 model at the cost of an AUC reduction of 0.02. \\
\textbf{Winograd algorithm.} When applied on dense convolution, Winograd reduces communication cost by an average of $2.12\times$ w.r.t. dense models without any reduction in accuracy.
For DenseNet-121, this provides a competitive alternative to efficient operators which marginally reduce AUC.\\
\textbf{Composition of Winograd with efficient operators.} Winograd algorithm can be applied to all model variants, as shown for the ShuffleNetV2 backbone. 
On the dense variant, the algorithm reduces communication by $1.7\times$ ($SDW$ vs $SD$).
However, for the three variants with efficient operators, the algorithm provides smaller gains between $1 - 2\%$. \\
\textbf{Latency.} For each variant, we also report the reduction in latency for a 2VM setup w.r.t. the baseline model.
We note that in secure 2-party computation, communication gains can be obtained at higher latency costs. 
However, with the optimizations proposed, gains in both metrics are observed. 

\noindent \textbf{Prior works.} 
As discussed, prior work has focused on optimizing non-linear layers which are not the bottleneck in CrypTFlow2.
Thus application of optimizations from prior work will only yield negligible reduction in communication costs relative to the large reduction we demonstrate through choices of models, operators, and algorithms.
For instance, even if we were to remove all ReLUs, as studied in \cite{delphi,cryptonas,safenet,deepreduce}, from DenseNet-121 (DD), the reduction in communication cost is only 0.8\%.

\section{Conclusion}
\label{sec:conclusion}
\noindent

We argue for designing models that have low 2PC costs to make secure inference practical.
To address this we propose combining 
(a) optimizations on the model such as using different network backbones, efficient clear-text operators, custom crypto-friendly operators, and fast convolution algorithms, with
(b) a secure inference framework like CrypTFlow2 which provides compilation of arbitrary deep models into functionally equivalent secure inference. 
In the chest X-ray interpretation task, we show that the optimizations lead to large reductions in secure inference costs with small drops in accuracy.

In the future,
we plan to apply orthogonal optimizations commonly used in secure inference such as varying bitwidths~\cite{coinn,sirnn}, reducing communication by increasing computation~\cite{caring,cheetah,silentot,ferret}, and hardware acceleration~\cite{piranha}.
We believe more gains remain to be unlocked at the intersection of design of secure inference frameworks and efficient models.

\end{document}